\documentclass{ws-ijmpd}
\usepackage{tipa}
\usepackage{amsmath}
\usepackage{txfonts}
\usepackage{stmaryrd}
\usepackage{amssymb}

\begin{document}

\markboth{Xu \& Guo}{Strange Matter: a state before black hole}

%
\catchline{}{}{}{}{}
%

\title{Strange Matter: a state before black hole}

\author{Renxin Xu\footnote{FAST Fellow (Distinguished Scholar)}~~ \and Yanjun Guo}

\address{School of Physics and KIAA,
         Peking University, Beijing 100871, P. R. China;\\
         {\tt r.x.xu@pku.edu.cn, guoyj10@pku.edu.cn}}

\maketitle


\begin{abstract}
Normal baryonic matter inside an evolved massive star can be intensely compressed by gravity after a supernova. General relativity predicts formation of a black hole if the core material is compressed into a singularity, but the real state of such compressed baryonic matter (CBM) before an event horizon of black hole appears is not yet well understood because of the non-perturbative nature of the fundamental strong interaction. Certainly, the rump left behind after a supernova explosion could manifest as a pulsar if its mass is less than the unknown maximum mass, $M_{\rm max}$. It is conjectured that pulsar-like compact stars are made of strange matter (i.e., with 3-flavour symmetry), where quarks are still localized as in the case of nuclear matter. In principle, different manifestations of pulsar-like objects could be explained in the regime of this conjecture. Besides compact stars, strange matter could also be manifested in the form of cosmic rays and even dark matter.
\end{abstract}


\vspace{0.5cm}

The baryonic part of the Universe is well-understood in the standard model of particle physics (consolidated enormously by the discovery of Higgs Boson), where quark masses are key parameters to make a judgment on the quark-flavour degrees of freedom at a certain energy scale.
Unlike the leptons, quarks could be described with mass parameters to be measured indirectly through their influence on hadronic properties since they are confined inside hadrons rather than free particles.
The masses of both up and down quarks are only a few MeV while the strange quark is a little bit heavier, with an averaged mass of up and down quarks, $m_{ud}=(3.40\pm0.25)$ MeV, as well as the strange quark mass of $m_s=(93.5\pm2.5)$ MeV obtained from lattice QCD (quantum chromo-dynamics) simulations.\cite{m_q}
For nuclei or nuclear matter, the separation between quarks is $\Delta x\sim 0.5$ fm, and the energy scale is then order of $E_{\rm nucl}\sim 400$ MeV according to Heisenberg's relation $\Delta x\cdot pc \sim \hbar c \simeq 200$ MeV$\cdot$fm.
One may then superficially understand why nuclei are of two (i.e., $u$ and $d$) flavours as these two flavours of quarks are the lightest.
However, because the nuclear energy scale is much larger than the mass differences between strange and up/down quarks, $E_{\rm nucl}\gg (m_s-m_{ud})$, why is the valence strangeness degree of freedom absolutely missing in stable nuclei?

We argue and explain in this paper that 3-flavour ($u$, $d$ and $s$) symmetry would be restored if the strong-interaction matter at low temperature is very big, with a length scale $\gg$ the electron Compton wavelength $\lambda_{\rm e}=h/(m_ec)\simeq 0.024$ \AA. We call this kind of matter as {\em strange matter} too, but it is worth noting that quarks are still localized with this definition (in analogy to 2-flavour symmetric nuclei) because the energy scale here (larger than but still around $E_{\rm nucl}$) is still much smaller than the perturbative scale of QCD dynamics, $\Lambda_\chi>1$ GeV.
We know that normal nuclei are relatively small, with length scale $(1\sim 10)$ fm $\ll \lambda_{\rm e}$, and it is very difficult for us to gather up huge numbers ($>10^9$) of nuclei together because of the Coulomb barrier between them in laboratory. Then, where could one find a large nucleus with possible 3-flavour symmetry (i.e., strange matter)?

Such kind of strange matter can only be created through extremely astrophysical events.
A good candidate of strange matter could be the supernova-produced rump left behind after core-collapsing of an evolved massive star, where normal {\em micro}-nuclei are intensely compressed by gravity to form a single {\em gigantic} nucleus (also called as compressed baryonic matter, CBM), the prototype of which was speculated and discussed firstly by Lev Landau.\cite{Landau1932}
The strange matter object could manifest the behaviors of pulsar-like compact stars if its mass is less than $M_{\rm max}$, the maximum mass being dependent on the equation of state of strange matter, but it could soon collapse further into a black hole if its mass $>M_{\rm max}$.
We may then conclude that strange matter could be the state of gravity-controlled CBM before an event horizon comes out (i.e., a black hole forms).

This paper is organized as follows.
In \S1, the gravity-compressed dense matter (a particular form of CBM), a topic relevant to Einstein's general relativity, is introduced in order to make sense of realistic CBM/strange matter in astrophysics.
We try to convince the reader that such kind of astrophysical CBM should be in a state of strange matter, which would be distinguished significantly from the previous version of strange quark matter, in \S2.
Cold strange matter would be in a solid state due to strong color interaction there, but the solution of a solid star with sufficient rigidity is still a challenge in general relativity. Nevertheless, the structure of solid strange star is presented (\S3) in the very simple case for static and spherically symmetric objects.
Different manifestations and astrophysical implications of strange matter are broadly discussed in \S4.
Finally, \S5 is a brief summary.

\section{Dense matter compressed by gravity}

As the first force recognized among the four fundamental interactions, gravity is mysterious and fascinating because of its unique feature.
Gravity is universal, which is well known from the epoch of Newton's theory.
Nothing could escape the control of gravity, from the falling of apple towards the Earth, to the motion of moon in the sky.
In Einstein's general relativity, gravity is related to the geometry of curved spacetime.
This beautiful and elegant idea significantly influences our world view.
Spacetime is curved by matter/energy, while the motion of object is along the ``straight'' line (geodesic) of the curved spacetime.
General relativity has passed all experimental tests up to now.
However, there is intrinsic conflict between quantum theory and general relativity.
Lots of efforts have been made to quantize gravity, but no success has been achieved yet.

Gravity is extremely weak compared to the other fundamental forces, so it is usually ignored in micro-physics.
Nonetheless, on the scale of universe, things are mostly controlled by gravity because it is long-range and has no screening effect.
One century has passed since Einstein established general relativity, but only a few solutions to the field equation have been found, among which three solutions are most famous and useful.
The most simple case is for static and spherical spacetime, and the solution was derived by Schwarzschild just one month after Einstein's field equation.
The Schwarzschild solution indicates also the existence of black hole, where everything is doomed to fall towards the center after passing through the event horizon.
Consider a non-vancum case with ideal fluid as source, the field equation could be transformed to Tolman-Oppenheimer-Vokoff equation\cite{TOV}, which could be applied to the interior of pulsar-like compact stars.
Based on the so-called cosmological Copernicus principle, Friedmann equation can be derived with the assumption of homogenous and isotropic universe, which sets the foundation of cosmology.
These three solutions of Einstein's field equation represent the most frontier topics in modern astrophysics.

At the late stage of stellar evolution, how does the core of massive star collapse to a black hole?
Or equivalently, how is normal baryonic matter squeezed into the singularity?
What's the state of compressed baryonic matter (CBM) before collapsing into a black hole?

We are focusing on these questions in this chapter.
In the standard model of particle physics, there are totally six flavours of quarks.
Among them, three ($u$, $d$ and $s$) are light, with masses $< 10^2$  MeV, while other three flavours ($c$, $t$ and $b$), with mass $> 10^3$ MeV, are too heavy to be excited in the nuclear energy scale, $E_{\rm nucl}\simeq 400$ MeV.
However, the ordinary matter in our world is built from $u$ and $d$ quarks only, and the numbers of these two flavours tend to be balanced in a stable nucleus.
It is then interesting to think philosophically about the fact that our baryonic matter is 2-flavour symmetric.

An explanation could be: micro-nuclei are too small to have 3-flavour symmetry, but bigger is different.
In fact, rational thinking about stable strangeness dates back to 1970s\cite{Bodmer1971}, in which Bodmer speculated that so-called ``collapsed nuclei'' with strangeness could be energetically favored if baryon number $A>A_{\min}$, but without quantitative estimation of the minimal number $A_{\min}$.
Bulk matter composed of almost {\em free} quarks ($u$, $d$, and $s$) was then focused on\cite{Itoh1970,Witten1984}, even for astrophysical manifestations\cite{Alcock1986,Haensel1986}.
CBM can manifest as a pulsar if the mass is not large enough to form a black hole, and strange quark matter could possibly exist in compact stars,
either in the core of neutron star (i.e., mixed or hybrid stars\cite{Ivanenko1969}) or as the whole star (strange quark star\cite{Itoh1970,Alcock1986,Haensel1986}).
Although the asymptotic freedom is well recognized, one essential point is whether the color coupling between quarks is still perturbative in astrophysical CBM so that quarks are itinerant there.
In case of non-perturbative coupling, the strong force there might render quarks grouped in so-called {\em quark-clusters}, forming a nucleus-like strange object\cite{Xu03} with 3-flavour symmetry, when CBM is big enough that relativistic electrons are inside (i.e., $A>A_{\rm min}\simeq 10^9$).
Anyway, we could simply call 3-flavour baryonic matter as {\em strange matter}, in which the constituent quarks could be either itinerant or localized.

Why is big CBM strange? This is actually a conjecture to be extensively discussed in the next section, but it could be reasonable.

\section{A Bodmer-Witten's conjecture generalized}


Besides being meaningful for understanding the nature of sub-nucleon at a deeper level, strangeness would also have a consequence of the physics of super-dense matter.
The discovery of strangeness (a general introduction to Murray Gell-Mann and his strangeness could be found in the biography by Johnson\cite{Johnson1999}) is known as a milestone in particle physics since our normal baryons are non-strange.
Nonetheless, condensed matter with strangeness should be worth exploring as the energy scale $E_{\rm nucl}\gg m_s$ at the nuclear and even supra-nuclear densities.

Previously, bulk strange object (suggested to be 3-flavour {\em quark} matter) is speculated to be the absolutely stable ground state of strong-interaction matter, which is known as the Bodmer-Witten's conjecture.\cite{Bodmer1971,Witten1984}
But we are discussing a general conjecture in next subsections, arguing that quarks might not be necessarily free in stable strange matter, and would still be hadron-like localized as in a nucleus if non-perturbative QCD effects are significant (i.e., $E_{\rm nucl}<\Lambda_\chi$) and the repulsive core keeps to work in both cases of 2-flavour (non-strange) nuclear matter and 3-flavour (strange) matter.
In this sense, protons and neutrons are of 2-flavour quark clusters, while strange matter could be condensed matter of quark clusters with strangeness (i.e., strange quark-clusters).

Summarily, it is well known that micro-nuclei are non-strange, but macro-nuclei in the form of CBM could be strange.
Therefore, astrophysical CBM and nucleus could be very similar, but only with a simple change from non-strange to strange: ``2'' $\rightarrow$ ``3''.
We are explaining two approaches to this strange quark-cluster matter state, bottom-up and top-down, respectively as following.

\subsection{Macro-nuclei with 3-flavour symmetry: bottom-up}

Micro-nucleus is made up of protons and neutrons, and there is an observed tendency to have equal numbers of protons ($Z$) and
neutrons ($N$).
In liquid-drop model, the mass formula of a nucleus with atomic number $A (= Z + N)$ consists of five terms,
\begin{equation}
E(Z, N)= a_{\rm v}A - a_{\rm s}A^{2/3} - a_{\rm sym} \frac{(N-Z)^2}{A} - a_{\rm c} \frac{Z(Z-1)}{A^{1/3}} + a_{\rm p}\frac{\Delta(N,Z)}{A^{1/2}},
\end{equation}
where the third term is for the symmetry energy, which vanishes with equal number of protons and neutrons.
This nuclear symmetry energy represents a symmetry between proton and neutron in the nucleon degree of freedom, and is actually that of $u$ and $d$ quarks in the quark degree\cite{Li_Chen2008}.
The underlying physics of symmetry energy is not well understood yet.
If the nucleons are treated as Fermi gas, there is a term with the same form as symmetry energy in the formula of Fermi energy, known as the kinetic term of nuclear symmetry energy.
But the interaction is not negligible, and the potential term of symmetry energy would be significant.
Recent scattering experiments show that, because of short-range interactions, the neutron-proton pairs are nearly 20 times as prevalent as proton-proton (and neutron-neutron by inference) pairs,\cite{Subedi2008,Hen2014} which hints that the potential term would dominate in the symmetry energy.

Since the electric charges of $u$ and $d$ quarks is +2/3 are -1/3 respectively, 2-flavour symmetric strong-interaction matter should be positively charged, and electrons are needed to keep electric neutrality.
The possibility of electrons inside a nucleus is negligible because the nuclear radius is much smaller than the Compton wavelength $\lambda_{\rm e} \sim 10^3 \, {\rm fm}$, and the lepton degree of freedom would then be not significant for nucleus.
Therefore, electrons contribute negligible energy for micro-nuclei, as the coupling constant of electromagnetic interaction ($\alpha_{\rm em}$) is much less than that of strong interaction ($\alpha_{\rm s}$).
The kinematic motion of electrons is bound by electromagnetic interaction, so $p^2/m_{\rm e} \sim e^2/l$.
From Heisenberg's relation, $p \cdot l \sim \hbar$,
combining the above two equations, we have $l \sim \frac{1}{\alpha_{\rm em}} \frac{\hbar c}{m_{\rm e}c^2}$, and the interaction energy is order of $e^2/l\sim \alpha^2_{\rm em}m_{\rm e}c^2 \sim 10^{-5}\, {\rm MeV}$.

However, bigger is different, and there might  be 3-flavour symmetry in gigantic/macro-nuclei, as  electrons are inside a gigantic nucleus.
With the number of nucleons $A > 10^9$, the scale of macro-nuclei should be larger than the Compton wavelength of electrons $\lambda_{\rm e} \sim 10^3 \, {\rm fm}$.
If 2-flavour symmetry keeps, macro-nucleus will become a huge Thomson atom with electrons evenly distributed there.
Though Coulomb energy could not be significant, the Fermi energy of electrons are not negligible, to be $E_{\rm F}\sim \hbar c n^{1/3} \sim 10^2 \, {\rm MeV}$.
However, the situation becomes different if strangeness is included: no electrons exist if the matter is composed by equal numbers of
light quarks of $u$, $d$, and $s$ in chemical equilibrium.
In this case, the 3-flavor symmetry, an analogy of the symmetry of
$u$ and $d$ in nucleus, may results in a ground state of matter for
gigantic nuclei. Certainly the mass difference between $u$/$d$ and $s$ quarks would also break the 3-flavour symmetry, but the interaction between quarks could lower the effect of mass differences and favor the restoration of 3-flavour symmetry.
If macro-nuclei are almost 3-flavour symmetric, the contribution of electrons would be negligible, with $n_{\rm e} \ll n_{\rm q}$ and $E_{\rm F} \sim 10 \, {\rm MeV}$.

The new degree of freedom (strangeness) is also possible to be excited, according to an order-of-magnitude estimation from either Heisenberg's relation (localized quarks) or Fermi energy (free quarks).
For quarks localized with length scale $l$, from Heisenberg's uncertainty relation,
the kinetic energy would be of $\sim p^2/m_{\rm q} \sim \hbar^2/(m_{\rm q} l^2)$,
which has to be comparable to the color interaction energy of $E\sim \alpha_{\rm s} \hbar c/l$ in order to have a bound state, with an assumption of Coulomb-like strong interaction.
One can then have if quarks are dressed,
\begin{equation}\label{}
 l \sim \frac{1}{\alpha_{\rm s}} \frac{\hbar c}{m_{\rm q} c^2} \simeq \frac{1}{\alpha_{\rm s}} \, {\rm fm, }~~
 E \sim \alpha_{\rm s}^2 m_{\rm q} c^2 \simeq 300 \alpha_{\rm s}^2 \, {\rm MeV.}
\end{equation}
As $\alpha_{\rm s}$ may well be close to and even greater than 1 at several times the
nuclear density, the energy scale would be approaching and even larger than
$\sim$400 MeV.
A further calculation of Fermi energy also gives,
\begin{equation}\label{}
  E_{\rm F}^{\rm NR} \approx \frac{\hbar^2}{2m_{\rm q}} (3\pi^2)^{2/3} \cdot n^{2/3} =380 \, {\rm MeV},
\end{equation}
if quarks are considered moving non-relativistically, or
\begin{equation}\label{}
  E_{\rm F}^{\rm ER} \approx \hbar c (3\pi^2)^{1/3} \cdot n^{1/3} = 480 \, {\rm MeV},
\end{equation}
if quarks are considered moving extremely relativistically.
So we have the energy scale $\sim 400 \, {\rm MeV}$ by either Heisenberg's relation or Fermi energy, which could certainly be larger the mass difference ($\sim 100\, {\rm MeV}$) between $s$ and $u/d$ quarks,
However, for micro-nuclei where electron contributes also negligible energy, there could be 2-flavour (rather than 3-flavour) symmetry because $s$ quark mass is larger than $u/d$ quark masses.
We now understand that it is more economical to have 2-flavour micro-nuclei because of massive $s$-quark and negligible electron kinematic energy, whereas macro/gigantic-nuclei might be 3-flavour symmetric.

The 2-flavour micro-nucleus consists of $u$ and $d$ quarks grouped in nucleons, while the 3-flavour macro-nucleus is made up of $u$, $d$ and $s$ quarks grouped in so-called strange quark-clusters.
Such macro-nucleus with 3-flavour symmetry can be named as strange quark-cluster matter, or simply strange matter.

\subsection{Macro-nuclei with 3-flavour symmetry: top-down}

Besides this bottom-up scenario (an approach from the hadronic state), we could also start from deconfined quark state with the inclusion of stronger and stronger interaction between quarks (a top-down scenario).

The underlying theory of the elementary strong interaction is believed to be quantum chromodynamics (QCD), a non-Abelian {\em SU}(3) gauge theory.
In QCD, the effective coupling between quarks decreases with energy (the {\em asymptotic freedom})\cite{Gross1973,Politzer1973}.
Quark matter (or quark-gluon plasma), the soup of deconfined quarks and gluons, is a direct consequence of asymptotic freedom when temperature or baryon density are extremely enough.
Hot quark matter could be reproduced in the experiments of relativistic heavy ion collisions.
Ultra-high chemical potential is required to create cold quark matter, and it can only exist in rare astrophysical conditions, the compact stars.

What kind of cold matter can we expect from QCD theory, in effective
models, or even based on phenomenology?
This is a question too hard to answer because of (i) the
non-perturbative effects of strong interaction between quarks at low
energy scales and (ii) the many-body problem due to vast assemblies
of interacting particles.
A color-superconductivity (CSC) state is focused on in
QCD-based models, as well as in phenomenological ones\cite{csc08}.
The ground state of extremely dense quark matter could certainly be
that of an ideal Fermi gas at an extremely high density. Nevertheless, it has been found that the
highly degenerate Fermi surface could be unstable against the formation of
quark Cooper pairs, which condense near the Fermi surface due to the
existence of color-attractive channels between the quarks. A
BCS-like color superconductivity, similar to electric
superconductivity, has been formulated within perturbative QCD at
ultra-high baryon densities. It has been argued, based on QCD-like
effective models, that color superconductivity could also occur even
at the more realistic baryon densities of pulsar-like compact
stars\cite{csc08}.

Can the realistic stellar densities be high enough to justify the
use of perturbative QCD?
It is surely a challenge to calculate the coupling constant, $\alpha_s$, from first principles.
Nevertheless, there are
some approaches to the non-perturbative effects of QCD, one of which
uses the Dyson-Schwinger equations tried by Fischer et al.,\cite{dse1,dse2} who
formulated,
\begin{equation}
\alpha_s(x)={\alpha_s(0)\over \ln(e+a_1x^{a_2}+b_1x^{b_2})},
\label{dse}%
\end{equation}
where $a_1=5.292~{\rm GeV}^{-2a_2}$, $a_2=2.324$, $b_1=0.034~{\rm
GeV}^{-2b_2}$, $b_2=3.169$, $x=p^2$ with $p$ the typical momentum in
GeV, and that $\alpha_s$ freezes at $\alpha_s(0)=2.972$.
For our case of assumed dense quark matter at $\sim 3\rho_0$, the chemical
potential is $\sim 0.4$ GeV, and
then $p^2\simeq 0.16$ GeV$^2$. Thus, it appears that the coupling in
realistic dense quark matter should be greater than 2, being close
to 3 in the Fischer's estimate presented in Eq.(\ref{dse}).
Therefore, this surely means that a weakly coupling treatment could be
dangerous for realistic cold quark matter (the interaction energy $\sim 300\alpha_s^2$ MeV could even be much larger than the Fermi energy), i.e., the non-perturbative effect
in QCD should not be negligible if we try to know the real state of
compact stars.
It is also worth noting that the dimensionless electromagnetic coupling
constant (i.e., the fine-structure constant) is $1/137<0.01$, which
makes QED tractable.
That is to say, a weakly coupling strength comparable with that of
QED is possible in QCD only if the density is unbelievably and
unrealistically high ($n_{\rm B}>10^{123}n_0$! with $n_0=0.16$ fm$^{-3}$ the baryon density of nuclear matter).

Quark-clusters may form in relatively low temperature
quark matter due to the strong interaction (i.e., large $\alpha_s$), and the clusters could locate in periodic lattices (normal solid) when
temperature becomes sufficiently low.
Although it is hitherto impossible to know if quark-clusters could
form in cold quark matter via calculation from first principles,
there could be a few points that favor clustering.
{\em Experimentally}, though quark matter is argued to be weakly
coupled at high energy and thus deconfined, it is worth noting
that, as revealed by the recent achievements in relativistic heavy
ion collision experiments, the interaction between quarks in a
fireball of quarks and gluons is still very strong (i.e. the
strongly coupled quark-gluon plasma, sQGP\cite{Shuryak}).
The strong coupling between quarks may naturally render quarks
grouped in clusters, i.e., a condensation in position space rather
than in momentum space.
{\em Theoretically}, the baryon-like particles in quarkyonic
matter\cite{Quarkyonic07} might be grouped further due to residual
color interaction if the baryon density is not extremely high,
and quark-clusters would form then at only a few nuclear density.
Certainly, more elaborate research work is necessary.

For cold quark matter at $3n_0$ density, the distance between quarks
is $\sim$ fm $\gg$ the Planck scale $\sim 10^{-20} \, {\rm fm}$, so
quarks and electrons can well be approximated as point-like
particles.
If Q$_\alpha$-like clusters are created in the quark
matter\cite{Xu03}, the distance between clusters are $\sim 2 \, {\rm fm}$.
The length scale $l$ and color interaction energy of quark-clusters has been estimated by
the uncertainty relation, assuming quarks are as dressed (the constituent quark mass is
$m_{\rm q}\sim 300$ MeV) and move non-relativistically in a cluster.
We have $l\sim \hbar c/(\alpha_{\rm s} m_{\rm q} c^2)\simeq 1$ fm if $\alpha_{\rm s}\sim 1$, and the color interaction energy $\sim \alpha_{\rm s}^2 m_{\rm q} c^2$ could be greater than the baryon chemical potential if $\alpha_{\rm s} \gtrsim 1$.
The strong coupling could render quarks grouped in position space to form clusters, forming a nucleus-like strange object\cite{Xu03} with 3-flavour symmetry, if it is big enough that relativistic electrons are inside (i.e., $A>A_{\rm min}\simeq 10^9$).
Quark-clusters could be considered as classical particles in cold quark-cluster matter and would be in lattices at a lower temperature.

In conclusion, quark-clusters could emerge in cold dense matter because of the
strong coupling between quarks.
The quark-clustering phase has high density and the strong
interaction is still dominant, so it is different from the usual
hadron phase, and on the other hand, the quark-clustering phase is
also different from the conventional quark matter phase which is
composed of relativistic and weakly interacting quarks.
The quark-clustering phase could be considered as an intermediate
state between hadron phase and free-quark phase, with deconfined
quarks grouped into quark-clusters.

\subsection{Comparison of micro-nuclei and macro-nuclei}

In summary, there could be some similarities and differences between micro-nuclei and macro-nuclei, listed as following.

{\em Similarity} 1: Both micro-nuclei and macro-nuclei are self-bound by the strong color interaction, in which quarks are localized in groups called generally as quark-clusters. We are sure that there are two kinds of quark-clusters inside micro-nuclei, the proton (with structure {$uud$}) and neutron ({$udd$}), but don't know well the clusters in macro-nuclei due to the lack of detailed experiments related.

{\em Similarity} 2: Since the strong interaction might not be very sensitive to flavour, the interaction between general quark-clusters should be similar to that of nucleon, which is found to be Lennard-Jones-like by both experiment and modeling. Especially, one could then expect a hard core\cite{Wilczek2007} (or repulsive core) of the interaction potential between strange quark-clusters though no direct experiment now hints this existence.

{\em Difference} 1: The most crucial difference is the change of flavour degree of freedom, from two ($u$ and $d$) in micro-nuclei to three ($u$, $d$ and $s$) in macro/gigantic-nuclei. We could thus have following different aspects derived.

{\em Difference} 2: The number of quarks in a quark-cluster is 3 for micro nuclei, but could be 6, 9, 12, and even 18 for macro nuclei, since the interaction between $\Lambda$-particles could be attractive\cite{Beane2011,Inoue2011} so that no positive pressure can support a gravitational star of $\Lambda$-cluster matter. We therefore call proton/neutron as light quark-clusters, while the strange quark-cluster as heavy clusters because of (1) massive $s$-quark and (2) large number of quarks inside.

{\em Difference} 3: A micro nucleus could be considered as a quantum system so that one could apply quantum mean-field theory, whereas the heavy clusters in strange matter may be classical particles since the quantum wavelength of massive clusters might be even smaller than the mean distance between them.

{\em Difference} 4: The equation of state (EoS) of strange matter would be stiffer\cite{LGX2013} than that of nuclear matter because the clusters in former should be non-relativistic but relativistic in latter. The kinematic energy of a cluster in both micro- and macro-nuclei could be $\sim 0.5 \, {\rm GeV}$, which is much smaller than the rest mass (generally $\gtrsim 2 \, {\rm GeV}$) of a strange quark-cluster.

{\em Difference} 5: Condensed matter of strange quark-clusters could be in a solid state at low temperature much smaller than the interaction energy between clusters. We could then expect solid pulsars\cite{Xu03} in nature although an idea of solid nucleus was also addressed\cite{Bertsch1974} a long time ago.

\subsection{A general conjecture of flavour symmetry}

The 3-flavour symmetry may hint the nature of strong interaction at low-energy scale.
Let's tell a story of science fiction about flavour symmetry.
Our protagonist is a fairy who is an expert in QCD at high energy scale (i.e., perturbative QCD) but knows little about spectacular non-perturbative effects.
There is a conversation between the fairy and God about strong-interaction matter.

God: ``I know six flavours of quarks, but how many flavours could there exist in stable strong-interaction matter?''

Fairy: ``It depends ... how dense is the matter? (\underline{aside}: the nuclear saturation density arises from the short-distance repulsive core, a consequence of non-perturbative QCD effect she may not know much.)''

God: ``Hum ... I am told that quark number density is about 0.48 fm$^{-3}$($3n_0$) there.''

Fairy: ``Ah, in this energy scale of $\sim 0.5$ GeV, there could only be light flavours (i.e., $u$, $d$ and $s$) in a stable matter if quarks are free.''

God: ``Two flavours ($u$ and $d$) or three flavours?''

Fairy: ``There could be two flavours of free quarks if strong-interaction matter is very small ($\ll \lambda_{\rm e}$), but would be three flavours for bulk strong-interaction matter (\underline{aside}: the Bodmer-Witten's conjecture).''

God: ``Small 2-flavour strong-interaction matter is very useful, and I can make life and mankind with huge numbers of the pieces. We can call them atoms.''

Fairy: ``Thanks, God! I can also help mankind to have better life.''

God: ``But ... are quarks really free there?''

Fairy: ``Hum ... there could be clustered quarks in both two and three flavour (\underline{aside}: a Bodmer-Witten's conjecture generalized) cases if the interaction between quarks is so strong that quarks are grouped together. You name a piece of small two flavour matter {\em atom}, what would we call a 3-flavour body in bulk?''

God: Oh ... simply, a {\em strange} object because of strangeness.

\section{Solid strange star in general relativity}

Cold strange matter with 3-flavour symmetry could be in a solid state because of (1) a relatively small quantum wave packet of quark-cluster (wave length $\lambda_{\rm q}<a$, where $a$ is the separation between quark-clusters) and (2) low temperature $T<(10^{-1}-10^{-2})U$ ($U$ is the interaction energy between quark-clusters). The packet scales $\lambda_{\rm q}\sim h/(m_qc)$ for free quark-cluster, with $m_q$ the rest mass of a quark cluster, but could be much smaller if it is constrained in a potential with depth of $>\hbar c/a\simeq 100$ MeV (2 fm$/a$).
A star made of strange matter would then be a solid star.

It is very fundamental to study static and spherically symmetric
gravitational sources in general relativity, especially for the
interior solutions.
The TOV solution\cite{TOV} is only for perfect
fluid. However, for solid strange stars, since the local press could
be {\em  anisotropic} in elastic matter, the radial pressure
gradient could be partially balanced by the tangential shear force
although a general understanding of relativistic, elastic bodies
has unfortunately not been achieved\cite{ks04}.
The origin of this local anisotropic force in solid quark stars
could be from the development of elastic energy as a star (i)
spins down (its ellipticity decreases) and (ii) cools (it may
shrink).
Release of the elastic as well as the gravitational energies would
be not negligible, and may have significant astrophysical
implications.

The structure of solid quark stars can be numerically calculated as following.
For the sack of simplicity, only
spherically symmetric sources are dealt with, in order to make sense of possible
astrophysical consequence of solid quark stars.
By introducing respectively radial and tangential pressures, $P$
and $P_\bot$, the stellar equilibrium equation of static
anisotropic matter in Newtonian gravity is\cite{hs97}: ${\rm d}P/{\rm d}r=-Gm(r)\rho/r^2+2(P_\bot-P)/r$,
where $\rho$ and $G$ denote mass density and the gravitational constant, respectively, and $m(r)=\int_0^r4\pi r^2\rho(x){\rm d}x$.
However, in Einstein's gravity, this equilibrium equation is
modified to be\cite{xty06},
\begin{eqnarray}
\frac{{\rm d}P}{{\rm d}r} = -\frac{Gm(r)\rho}{r^2}\frac{(1 +
\frac{P}{\rho c^2})(1 + \frac{4\pi r^3P}{m(r)c^2})}{1 -
\frac{2Gm(r)}{rc^2}} +
\frac{2\epsilon}{r}P,%
\label{P'}
\end{eqnarray}
where $P_\bot=(1+\epsilon)P$ is introduced. In case of isotropic
pressure, $\epsilon=0$, Eq.~(\ref{P'}) turns out to be the TOV
equation.
It is evident from Eq.~(\ref{P'}) that the radial pressure
gradient, $|{\rm d}P/{\rm d}r|$, decreases if $P_\bot>P$, which
may result in a higher maximum mass of compact stars.
One can also see that a sudden decrease of $P_\bot$ (equivalently of elastic force) in a star may
cause substantial energy release, since the star's radius
decreases and the absolute gravitational energy increases.

Starquakes may result in a sudden change of $\epsilon$, with an
energy release of the gravitational energy as well as the
tangential strain energy.
Generally, it is evident that the differences of radius,
gravitational energy, and moment of inertia increase
proportionally to stellar mass and the parameter $\epsilon$. This
means that an event should be more important for a bigger change
of $\epsilon$ in a quark star with higher mass.
Typical energy of $10^{44\sim 47} \, {\rm erg}$ is released during superflares of SGRs, and a giant starquake with $\epsilon \lesssim 10^{-4}$ could produce such a flare\cite{xty06}.
A sudden change of $\epsilon$ can also result in a jump of spin frequency, $\Delta \Omega/\Omega=-\Delta I/I$.
Glitches with $\Delta \Omega/\Omega\sim 10^{-10\sim -4}$ could occur for parameters of $M=(0.1\sim 1.4)M_\odot$ and $\epsilon=10^{-9\sim -4}$.
It is suggestive that a giantflare may accompany a high-amplitude glitch.

\section{Astrophysical manifestations of strange matter}

How to create macro-nuclei (even gigantic) in the Universe?
Besides a collapse event where normal baryonic matter is intensely compressed by gravity, strange matter could also be produced after cosmic hadronization.\cite{Witten1984}
Strange matter may manifest itself as a variety of objects with a broad mass spectrum, including compact objects, cosmic rays and even dark matter.


\subsection{Pulsar-like compact star: compressed baryonic matter after supernova}

In 1932, soon after Chandrasekhar found a unique mass (the mass
limit of white dwarfs), Landau speculated a state of matter, the
density of which ``becomes so great that atomic nuclei come in close
contact, forming one {\em gigantic nucleus}''\cite{Landau1932}.
A star composed mostly of such matter is called a ``neutron'' star,
and Baade and Zwicky even suggested in 1934 that neutron stars (NSs)
could be born after supernovae.
NSs theoretically predicted were finally {\em discovered} when
Hewish and his collaborators detected radio pulsars in 1967\cite{psr68}.
More kinds of pulsar-like stars, such as X-ray pulsars and X-ray
bursts in binary systems, were also discovered later, and all of
them are suggested to be NSs.

In gigantic nucleus, protons and electrons combined to form the neutronic state, which involves weak equilibrium between protons and neutrons.
However, the simple and beautiful idea proposed by Landau and others
had one flaw at least:
nucleons (neutrons and protons) are in fact {\em not} structureless
point-like particles although they were thought to be elementary
particles in 1930s.
A success in the classification of hadrons discovered in cosmic
rays and in accelerators leaded Gell-Mann to coin ``{\em
quark}'' with fraction charges ($\pm 1/3, \mp 2/3$) in
mathematical description, rather than in reality\cite{Gellmann1964}.
All
the six flavors of quarks ($u,d,c,s,t,b$) have experimental
evidence (the evidence for the last one, top quark, was reported
in 1995).
Is weak equilibrium among $u$, $d$ and $s$ quarks possible, instead of simply that between $u$ and $d$ quarks?

\begin{figure}
\centerline{\psfig{file=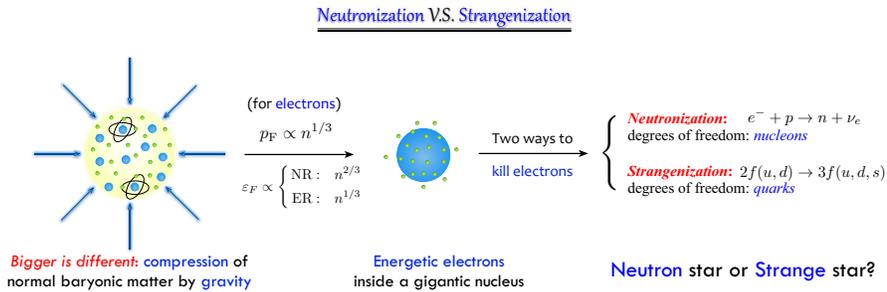,width=12cm}} \vspace*{8pt}
\caption{%
Neutronization and Strangenization are two competing ways to cancel energetic electrons.
\label{f1}}%
\end{figure}

At the late stage of stellar evolution, normal baryonic matter is intensely compressed by gravity in the core of massive star during supernova.
The Fermi energy of electrons are significant in CBM, and it is very essential to cancel the electrons by weak interaction in order to make lower energy state.
There are two ways to kill electrons as shown in Fig~\ref{f1}: one is via {\em neutronization}, $e^- + p \to n + \nu_{\rm e}$, where the fundamental degrees of freedom could be nucleons; the other is through {\em strangenization}, where the degrees of freedom are quarks.
While neutronization works for removing electrons, strangenization has both the advantages of minimizing the electron's contribution of kinetic energy and maximizing the flavour number, the later could be related to the flavour symmetry of strong-interaction matter.
These two ways to kill electrons are relevant to the nature of pulsar, to be neutron star or strange star, as summarized in Fig.~\ref{f1}.

\begin{figure}[!htb]
\centering
\includegraphics[width=10cm]{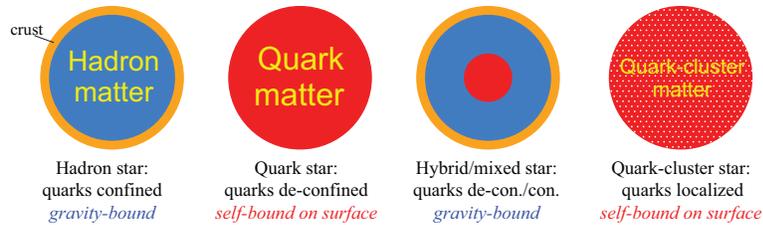}
\caption{Different models of pulsar's nature.
Hadron star and hybrid/mixed star are of conventional neutron stars, while
strangeness plays an important role for quark star and quark-cluster star (simply {\it strange star})
as a result of three light-flavour ({\it u}, {\it d} and {\it s}) symmetry.
\label{f2}}
\end{figure}

There are many speculations about the nature of pulsar due to unknown non-perturbative QCD at low energy.
Among different pulsar models, hadron star and hybrid/mixed star are conventional neutron stars, while quark star and quark-cluster star are strange stars with light flavour symmetry.
In hadron star model, quarks are confined in hadrons such as neutron/proton and hyperon, while a quark star is dominated by de-confined free quarks.
A hybrid/mixed star, with quark matter in its cores, is a mixture of hadronic and quark states.
However, a quark-cluster star, in which strong coupling causes individual quarks grouped in clusters, is neither a hadron star nor a quark star.
As an analog of neutrons, quark-clusters are bound states of several quarks, so to this point of view a quark-cluster star is more similar to a {\em real} giant nucleus of self-bound (not that of Landau), rather than a ``giant hadron'' which describes traditional quark stars.
Different models of pulsar's inner structure are illustrated in Fig.~\ref{f2}.

It is shown in Fig.~\ref{f2} that conventional neutron stars (hadron star and hybrid/mixed star) are gravity-bound, while strange stars (strange quark star and strange quark-cluster star) are self-bound on surface by strong force. This feature difference is very useful to identify observationally.
In the neutron star picture, the inner and outer cores and the crust keep chemical equilibrium at each boundary, so neutron star is bound by gravity.
The core should have a boundary and is in equilibrium with the ordinary matter because the star has a surface composed of ordinary matter.
There is, however, no clear observational evidence for a neutron star's surface, although most of authors still take it for granted that there should be ordinary matter on surface, and consequently a neutron star has different components from inner to outer parts.
Being similar to traditional quark stars, quark-cluster stars have almost the same composition from the center to the surface, and the quark matter surface could be natural for understanding some different observations.
It is also worth noting that, although composed of quark-clusters, quark-cluster stars are self-bound by the residual interaction between quark-clusters.
This is different from but similar to the traditional MIT bag scenario.
The interaction between quark-clusters could be strong enough to make condensed matter, and on the surface, the quark-clusters are just in the potential well of the interaction, leading to non-vanishing density but vanishing pressure.

Observations of pulsar-like compact stars, including surface and global properties, could provide hints for the state of CBM, as discussed in the following.

\subsubsection{Surface properties}

{\em Dirfting subpulses.} Although pulsar-like stars have many different manifestations, they are
populated by radio pulsars.
Among the magnetospheric emission models for pulsar radio radiative
process, the user-friendly nature of Ruderman-Sutherland\cite{RS1975}
model is a virtue not shared by others, and clear drifting
sub-pulses were first explained.
In the seminal paper, a vacuum gap was suggested above the polar cap of a pulsar.
The sparks produced by the inner-gap breakdown result in the subpulses, and the observed drifting feature is caused by ${\bf E \times B}$.
However, that model can only work in strict conditions for conventional neutron stars: strong
magnetic field and low temperature on surfaces of pulsars with ${\bf
\Omega \cdot B}<0$, while calculations showed, unfortunately, that
these conditions usually cannot be satisfied there.
The above model encounters the so-called ``binding energy
problem''. Calculations have shown that the binding energy of Fe at
the neutron star surface is $<1$ keV\cite{Fowlers1977,Lai2001}, which is not
sufficient to reproduce the vacuum gap.
These problems might be alleviated within a partially screened inner gap
model\cite{gm06} for NSs with ${\bf \Omega \cdot B}<0$, but could be
naturally solved for any ${\bf \Omega \cdot B}$ in the bare strange (quark-cluster) star
scenario.

The magnetospheric activity of bare quark-cluster star was investigated in quantitative details\cite{YX2011}.
Since quarks on the surface are confined by strong color interaction, the binding energy of quarks can be even considered as infinity compared to
electromagnetic interaction.
As for electrons on the surface, on one hand the potential barrier of the vacuum gap prevents electrons from streaming into the magnetosphere, on the other hand the total energy of electrons on the Fermi surface is none-zero.
Therefore, the binding energy of electrons is determined by the difference between the height of the potential barrier in the vacuum gap and the total energy of electrons.
Calculations have shown that the huge potential barrier built by the electric field in the vacuum gap above the polar cap can usually prevent electrons from streaming into the magnetosphere, unless the electric potential of a pulsar is sufficiently lower than that at the infinite interstellar medium.
In the bare quark-cluster star model, both positively and negatively charged particles on the surface are usually bound strongly enough to form a vacuum gap above its polar cap, and the drifting (even bi-drifting) subpulses can be understood naturally\cite{XQZ1999,QLZ04}.

{\em X-ray spectral lines.}
In conventional neutron star (NS)/crusted strange star models, an atmosphere exists above the surface of a central star.
Many theoretical calculations, first developed by Romani\cite{Romani1987}, predicted the existence of atomic features in the thermal X-ray emission of NS (also for crusted strange star) atmospheres, and advanced facilities of {\it Chandra} and {\it XMM-Newton} were then proposed to be constructed for detecting those lines.
One expects to know the chemical composition and magnetic field of the atmosphere through such observations, and eventually to constrain stellar mass and radius according to the redshift and pressure broadening of spectral lines.

However, unfortunately, none of the expected spectral features has been detected with certainty up to now, and this negative test may hint a fundamental weakness of the NS models.
Although conventional NS models cannot be completely ruled out by only non-atomic thermal spectra since modified NS atmospheric models with very strong surface magnetic fields\cite{HL03,TZD04} might reproduce a featureless spectrum too, a natural suggestion to understand the general observation could be that pulsars are actually bare strange (quark or quark-cluster) star\cite{Xu02}, almost without atoms there on the surfaces.

More observations, however, did show absorption lines of PSR-like stars, and the best absorption features were detected for the central compact object (CCO) 1E 1207.4-5209 in the center of supernova remnant PKS 1209-51/52, at $\sim0.7$ keV and $\sim1.4$ keV\cite{Sanwal2002,Mereghetti2002,Bignami2003}.
Although initially these features were thought to be due to atomic transitions of ionized helium in an atmosphere with a strong magnetic field, soon thereafter it was noted that these lines might be of electron-cyclotron origin, and 1E 1207 could be a bare strange star with surface field of $\sim 10^{11}$ G\cite{XWQ2003}.
Further observations of both spectra feature\cite{Bignami2003} and precise timing\cite{GH2007} favor the electron-cyclotron model of 1E 1207.

But this simple single particle approximation might not be reliable due to high electron density in strange stars, and Xu et al. investigated the global motion of the electron seas on the magnetized surfaces\cite{Xu2012}.
It is found that hydrodynamic surface fluctuations of the electron sea would be greatly affected by the magnetic field, and an analysis shows that the seas may undergo hydrocyclotron oscillations whose eigen frequencies are given by $\omega(l)=\omega_{c}/[l(l+1)]$, where
$l=1, 2, 3, ...$ and $\omega_{\rm c}=eB/mc$ is the cyclotron frequency.
The fact that the absorption feature of 1E 1207.4-5209 at $0.7$ keV is not much stronger than that at $1.4$ keV could be understood in this hydrocyclotron oscillations model, because these two lines with $l$ and $l+1$ could have nearly equal intensity, while the strength of the first harmonic is much smaller than that of the fundamental in the electron-cyclotron model.
Besides the absorption in 1E 1207.4-5209, the detected lines around $(17.5, 11.2, 7.5, 5.0)$ keV in the burst spectrum of SGR 1806-20 and those in other dead pulsars (e.g., radio quiet compact objects) would also be of hydrocyclotron origin\cite{Xu2012}.

{\em Planck-like continue spectra.}
The X-ray spectra from some sources (e.g., RX J1856) are well fitted by blackbody, especially with high-energy tails surprisingly close to Wien's formula: decreasing exponentially ($\propto e^{-\nu}$).
Because there is an atmosphere above the surface of neutron star/crusted strange stars, the spectrum determined by the radiative transfer in atmosphere should differ substantially from Planck-like one, depending on the chemical composition, magnetic field, etc.\cite{Zavlin1996}
Can the thermal spectrum of quark-cluster star be well described by Planck's
radiation law?
In bag models where quarks are nonlocal, one limitation is that bare
strange stars are generally supposed to be poor radiators in thermal X-ray
as a result of their high plasma frequency, $\sim 10$ MeV.
Nonetheless, if quarks are localized to form quark-clusters in cold
quark matter due to very strong interactions, a regular lattice of
the clusters (i.e., similar to a classical {\em solid} state)
emerges as a consequence of the residual interaction between
clusters\cite{Xu03}.
In this latter case, the metal-like solid quark matter would induce
a metal-like radiative spectrum, with which the observed thermal
X-ray data of RX J1856 can be fitted\cite{zxz04}.
Alternatively, other radiative mechanism in the electrosphere (e.g.,
electron bremsstrahlung in the strong electric
field\cite{Zakharov10} and even of negligible ions above the sharp surface) may also reproduce a Planck-like continue spectrum.

{\em Supernova and gamma-ray bursts. }
It is well known that the radiation fireballs of gamma-ray bursts (GRBs) and supernovae as a whole move towards the observer with a high Lorentz factor.\cite{Paczynski1986}
The bulk Lorentz factor of the ultrarelativistic fireball of GRBs is estimated to be order\cite{Meszaros1998} of $\Gamma \sim 10^2-10^3$.
For such an ultra-relativistic fireball, the total mass of baryons can not be too high, otherwise baryons would carry out too much energy of the central engine, the so-called ``baryon contamination''.
For conventional neutron stars as the central engine, the number of baryons loaded with the fireball is unlikely to be small, since neutron stars are gravity-confined and the luminosity of fireball is extremely high.
However, the baryon contamination problem can be solved naturally if the central compact objects are strange quark-cluster stars.
The bare and chromatically confined surface of quark-cluster stars separates baryonic matter from the photon and lepton dominated fireball.
Inside the star, baryons are in quark-cluster phase and can not escape due to strong color interaction, but $e^\pm$-pairs, photons and neutrino pairs can escape from the surface.
Thus, the surface of quark-cluster stars automatically generates a low baryon condition
for GRBs as well as supernovae.\cite{Ouyed05,Paczynski2005,Cheng1996}

It is still an unsolved problem to simulate supernovae successfully in the neutrino-driven explosion models of neutron stars.
Nevertheless, in the quark-cluster star scenario, the {\em bare} quark surfaces could be essential for successful explosions of both core and accretion-induced collapses\cite{xu05mn}.
A nascent quark-cluster star born in the center of GRB or supernova would radiate thermal emission due to its ultrahigh surface temperature\cite{Haensel1991}, and the photon luminosity is not constrained by the Eddington limit since the surface of quark-cluster stars could be bare and chromatically confined.
Therefore, in this photon-driven scenario,\cite{CYX07} the strong radiation pressure caused by enormous thermal emissions from quark-cluster stars might play an important role in promoting core-collapse supernovae.
Calculations have shown that the radiation pressure due to such strong thermal emission can push the overlying mantle away through photon-electron scattering with energy as much as $\sim10^{51}$ ergs.
Such photon-driven mechanism in core-collapse supernovae by forming a quark-cluster star inside the collapsing core is promising to
alleviate the current difficulty in core-collapse supernovae. The recent discovery of highly super-luminous supernova ASASSN-15lh, with a total observed energy $(1.1\pm 0.2)\times 10^{52}$ ergs,\cite{Dong16} might also be understood in this regime if a very massive strange quark-cluster star, with mass smaller than but approaching $M_{\rm max}$, forms.

\subsubsection{Global properties.}

{\em Free or torque-induced precession. }
Rigid body precesses naturally when spinning, either freely or by torque, but fluid one can hardly.
The observation of possible precession or even free precession of B1821-11\cite{Stairs00} and others could suggest a global solid structure for pulsar-like stars.
Low-mass quark stars with masses of $\lesssim 10^{-2}M_\odot$ and radii of a few kilometers are gravitationally force-free, and their surfaces could then be irregular (i.e., asteroid-like).
Therefore, free or torque-induced precession may easily be excited and expected with larger amplitude in low-mass quark stars.
The masses of AXPs/SGRs (anomalous X-ray pulsars/soft gamma-ray repeaters) could be approaching the mass-limit ($>1.5 M_\odot$)
in the AIQ (accretion-induced quake) model\cite{Xu07b}; these objects could then manifest no or weak
precession as observed, though they are more likely than CCOs/DTNs
(eg., RX J1856) to be surrounded by dust disks because of their
higher masses (thus stronger gravity).

{\em Normal and slow glitches.}
A big disadvantage that one believes that pulsars are strange quark stars lies in the fact that the observation of pulsar glitches conflicts with the hypothesis of conventional quark stars in fluid states\cite{Alpar87,BHV90} (e.g., in MIT bag models).
That problem could be solved in a solid quark-cluster star model since a solid stellar object would inevitably result in star-quakes when strain
energy develops to a critical value.
Huge energy should be released, and thus large spin-change occurs, after a quake of a solid quark star.
Star-quakes could then be a simple and intuitional mechanism for pulsars to have glitches frequently with large amplitudes.
In the regime of solid quark star, by extending the model for normal glitches\cite{z04}, one can also model pulsar's slow glitches\cite{px07} not  well understood in NS models.
In addition, both types of glitches without (Vela-like, Type I) and  with (AXP/SGR-like, Type II) X-ray enhancement could be naturally understood in the star-quake model of solid strange star,\cite{zhou14} since the energy release during a type I (for fast rotators) and a type II (for slow rotators) starquake are very different.

{\em Energy budget}.
The substantial free energy released after star-quakes, both elastic and gravitational, would power some extreme events detected in AXPs/SGRs and during GRBs.
Besides persistent pulsed X-ray emission with luminosity well in excess of the spin-down power, AXPs/SGRs show occasional bursts (associated possibly with glitches), even superflares with isotropic energy $\sim 10^{44-46}$ erg and initial peak luminosity $\sim 10^{6-9}$ times of the Eddington one.
They are speculated to be magnetars, with the energy reservoir of magnetic fields $\gtrsim 10^{14}$ G (to be still a matter of debate about the origin since the dynamo action might not be so effective and the strong magnetic field could decay effectively), but failed predictions are challenges to the model.\cite{tx11}
However, AXPs/SGRs could also be solid quark stars with surface magnetic fields similar to that of radio pulsars.
Star-quakes are responsible to both bursts/flares and glitches in the latter scenario,\cite{Xu07b} and kinematic oscillation energy could effectively power the magnetospheric activity.\cite{lxz15}

The most conspicuous asteroseismic manifestion of solid phase of quark stars is their capability of sustaining torsional shear oscillations induced by SGR's starquake\cite{Sergey09}.
In addition, there are more and more authors who are trying to connect the GRB central engines to SGRs' flares, in order to understand different GRB light-curves observed, especially the internal-plateau X-ray emission.\cite{xl09,dlx11}

{\em Mass and radius of compact star}. The EoS of quark-cluster matter would be stiffer than that of nuclear matter, because (1) quark-cluster should be non-relativistic particle for its large mass, and (2) there could be strong short-distance repulsion between quark-clusters.
Besides, both the problems of hyperon puzzle and quark-confinement do not exist in quark-cluster star.
Stiff EoS implies high maximum mass, while low mass is a direct consequence of self-bound surface.

It has been addressed that quark-cluster stars could have high maximum masses ($>2M_\odot$) as well as very low masses ($<10^{-2}M_\odot$).\cite{LX09}
Later radio observations of PSR J1614-2230, a binary
millisecond pulsar with a strong Shapiro delay signature, imply that
the pulsar mass is 1.97$\pm$0.04 $M_\odot$\cite{Demorest12}, which
indicates a stiff EoS for CBM.
Another 2$M_\odot$ pulsar is also discovered afterwards\cite{Antoniadis13}.
It is conventionally thought that the state of dense matter softens and thus
cannot result in high maximum mass if pulsars are quark stars, and that the
discovery of massive 2$M_\odot$ pulsar may make pulsars unlikely to be quark stars.
However, quark-cluster star could not be ruled out by massive pulsars, and the observations
of pulsars with higher mass (e.g. $>2.5M_\odot$) , would even be a strong support to
quark-cluster star model, and give further constraints to the parameters.
The mass and radius of 4U 1746-37 could be constrained by PRE (photospheric radius expansion) bursts, on the assumption that the touchdown flux corresponds to Eddington luminosity and the obscure effect is included.\cite{Li2015}
It turns out that 4U 1746-37 could be a strange star with small radius.

There could be other observational hints of low-mass strange stars.
Thermal radiation components from some PSR-like stars are detected,
the radii of which are usually much smaller than 10 km in blackbody
models where one fits spectral data by Planck
spectrum,\cite{pavlov04} and Pavlov and Luna\cite{PL09} find
no pulsations with periods longer than $\sim 0.68$ s in the CCO of Cas A, and constrain stellar radius and mass
to be $R=(4\sim 5.5) \, {\rm km}$ and $M\lesssim 0.8M_\odot$ in hydrogen NS
atmosphere models.
Two kinds of efforts are made toward an understanding of the fact in
conventional NS models.
(1) The emissivity of NS's surface isn't simply of
blackbody or of hydrogen-like atmospheres.
The CCO in Cas A is
suggested to covered by a carbon atmosphere\cite{HH09}.
However, the spectra from some sources (e.g., RX J1856) are still puzzling, being
well fitted by blackbody.
(2) The small emission areas would represent
hot spots on NS's surfaces, i.e., to fit the X-ray spectra with at
least two blackbodies, but this has three points of weakness in NS
models.
$a$, about $P$ and $\dot P$. No or very weak pulsation has been detected in
some of thermal component-dominated sources (e.g., the Cas A
CCO\cite{PL09}), and the inferred magnetic field from $\dot P$ seems
not to be consistent with the atmosphere models at least for RX
J1856\cite{KK08}.
$b$, fitting of thermal X-ray spectra (e.g., PSR
J1852+0040) with two blackbodies finds two small emitting radii
(significantly smaller than 10 km), which are not yet
understood\cite{HG10}.
$c$, the blackbody temperature of the entire
surface of some PSR-like stars are much lower than those predicted
by the standard NS cooling models,\cite{LiXH05} even provided that
hot spots exist.
Nevertheless, besides that two above, a {\em natural} idea could be
that the detected small thermal regions ({\em if} being global) of
CCOs and others may reflect their small radii (and thus low masses
in quark-cluster star scenario).\cite{xu05mn}

\begin{figure}[!htb]
\centering
\includegraphics[width=8cm]{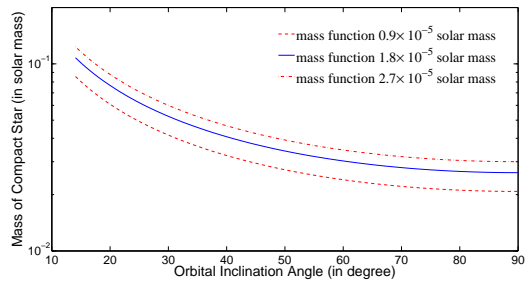}
\caption{%
The compact star mass as a function of orbital inclination for different values of mass function.
\label{f3}}
\end{figure}

Another low-mass strange (quark-cluster) star could be 4U 1700+24.
Because of strangeness barrier existing above a quark-cluster surface, a strange star may be surrounded by a hot corona or an atmosphere, or even a crust for different accretion rates.
Both the redshifted O VIII Ly-$\alpha$ emission line (only $z=0.009$) and the change in the blackbody radiation area (with an inferred scale of $\sim (10-10^2)$ m) could naturally be understood if 4U 1700+24 is a low-mass quark-cluster star which exhibits weak wind accretion.\cite{1700}
Additionally, the mass function via observing the G-type red giant company is only $f_o=(1.8\pm 0.9)\times 10^{-5}M_\odot$,\cite{Galloway02} from which the derived mass of compact star should be much lower than $1 M_\odot$ unless there is geometrically fine-tuning (inclination angle $i<2^{\rm o}$, see Fig.~\ref{f3}).
All these three independent observations (redshift, hot spot and mass function) may point to the fact that 4U 1700+24 could be a low mass strange quark-cluster star.

Future observations with more advanced facilities, such as FAST and SKA, could provide more observational hints for the nature of CBM.
Pulsar mass measurement could help us find more massive pulsar, while measurement of the momentum of inertia may give information on the radius.
Searching sub-millisecond pulsars could be an expected way to
provide clear evidence for (low-mass) quark stars.
Normal neutron stars can {\rm not} spin with periods less
than $\sim 0.5M_1^{1/2}R_6^{-3/2} \, {\rm ms}$ ($R_6=R/10^6\, {\rm cm}$), as the rotation is limited by Kepler frequency.
But low-mass bare strange stars has no such limitation on the spin period, which could be even less than 1 ms.
We need thus a much short sampling time, and would deal with then
a huge amount of data in order to find a sub-millisecond pulsars.
Besides, the pulse profile of pulsar is helpful for the understanding of its magnetospheric activity.

\subsection{Strange matter in cosmic rays and as dark matter candidate}

Strange quark-nuggets, in the form of cosmic rays, could be ejected during the birth of central compact star,\cite{XW03} or during collision of strange stars in a binary system spiraling towards each other due to loss of orbital energy via gravitational waves.\cite{Madsen05}

A strangelet with mass per baryon $< 940$ MeV (i.e., binding energy
per baryon $\gtrsim 100$ MeV) could be stable in cosmic rays, and
would decay finally into nucleons when collision-induced decrease of
baryon number make it unstable due to the increase of surface
energy.
When a stable strangelet bombards the atmosphere of the Earth, its fragmented nuggets
may decay quickly into $\Lambda$-particles by strong interaction and further
into nucleons by weak interaction.
What if a strange nugget made of quark clusters bombards the Earth? It is interesting and necessary to investigate.

In the early Universe ($\sim 10 \, {\rm \mu s}$), quark-gluon plasma condenses to form hadron gas during the QCD phase transition.
If the cosmological QCD transition is first-order, bubbles of hadron gas are nucleated and grow until they merge and fill up the whole Universe.
A separation of phases during the coexistence of the hadronic and the quark phases could gather a large number of baryons in strange nuggets.\cite{Witten1984}
If quark clustering occurs, evaporation and boiling may be suppressed, and strange nuggets may survive and contribute to the dark matter today.
Strange nuggets as cold quark matter may favor the formation of seed black holes in primordial halos, alleviating the current difficulty of quasars at redshift as high as $z \sim 6$,\cite{LX10} and the small pulsar glitches detected may hint the role of strange nuggets.\cite{LX16}

\section{Conclusions}

Although normal micro-nuclei are 2-flavour symmetric, we argue that 3-flavour symmetry would be restored in macro/gigantic-nuclei compressed by gravity during a supernova.
Strange matter is conjectured to be condensed matter of 3-flavour quark-clusters, and future advanced facilities (e.g., FAST, SKA) would provide clear evidence for strange stars.
Strange nuggets manifested in the form of cosmic rays and even dark matter have significant astrophysical consequences, to be tested observationally.

\vspace{3mm}
\noindent%
{\bf Acknowledgements.}
This work is supported by the National Basic Research Program of China
(No. 2012CB821801) and NNSFC (No. 11225314).
The FAST FELLOWSHIP is supported by the Special Funding for Advanced Users,
budgeted and administrated by Center for Astronomical Mega-Science, Chinese
Academy of Sciences (CAS).
We would like to thank Ms. Yong Su for reading and checking \S2.4.


\begin{thebibliography}{0}

\bibitem{m_q}
K. A. Olive et al. (Particle Data Group), Chin. Phys. {\bf C38} (2014) 090001.

\bibitem{Landau1932}
L. Landau, Phys. Z. Sowjetunion {\bf 1} (1932) 285.

\bibitem{TOV}
J. R. Oppenheimer and G. B. Volkoff, Physical Review {\bf 55} (1939) 374.

\bibitem{Bodmer1971}
A. R. Bodmer, Physical Review {\bf D4} (1971) 16.

\bibitem{Itoh1970}
N. Itoh,  Prog. Theor. Phys. {\bf 44} (1970) 291.

\bibitem{Witten1984}
E. Witten, Phys. Rev. D {\bf 30} (1984) 272.

\bibitem{Alcock1986}
C. Alcock, E. Farhi and A. Olinto, ApJ {\bf 310} (1986) 261.

\bibitem{Haensel1986}
P. Haensel, J. L. Zdunik and R. Schaeffer, A\&A {\bf 160} (1986) 121.

\bibitem{Ivanenko1969}
D. Ivanenko, D. F. Kurdgelaidze, Lett. Nuovo Cimento {\bf 2} (1969) 13.

\bibitem{Xu03}
R. X. Xu, ApJ {\bf 596} (2003)  L59.

\bibitem{Johnson1999}
G. Johnson, {\em Strange Beauty}, A Division of Random House, Inc., New York, 1999.

\bibitem{Li_Chen2008}
 B.~A.~Li, L.~W.~Chen and C.~M.~Ko, Physics Reports {\bf 464} (2008) 113.

\bibitem{Subedi2008}
R. Subedi et al., Science {\bf 320} (2008) 1476.

\bibitem{Hen2014}
O. Hen et al., Science {\bf 346} (2014) 614.

\bibitem{Gross1973}
D. J. Gross and F. Wilczek, Phys. Rev. Lett. {\bf 30} (1973) 1343.

\bibitem{Politzer1973}
H. D. Politzer, Phys. Rev. Lett. {\bf 30} (1973) 1346.

\bibitem{csc08}
M. J. Alford {\em et al.}, Rev. Mod. Phys. {\bf 80} (2008) 1455.

\bibitem{dse1}
C. S. Fischer, R. Alkofer {\em Phys. Lett.}, {\bf B536} (2002) 177.

\bibitem{dse2}
C. S. Fischer {\em J. Phys. G: Part. Nucl. Phys}, {\bf 32} (2006) R253.

\bibitem{Shuryak}
E. V. Shuryak, Prog. Part. \& Nucl. Phys. {\bf 62} (2009) 48.

\bibitem{Quarkyonic07}
L. McLerran, R. D. Pisarski, Nucl. Phys. {\bf A796} (2007) 83.

\bibitem{Wilczek2007}
F. Wilczek, Nature {\bf 445} (2007) 156.

\bibitem{Beane2011}
S. R. Beane, E. Chang, W. Detmold et al., Phys. Rev. Lett. {\bf 106} (2011) 162001.

\bibitem{Inoue2011}
T. Inoue {\it et al.}, Phys. Rev. Lett. {\bf 106} (2011) 162002.

\bibitem{LGX2013}
X. Y. Lai, C. Y. Gao and R. X. Xu, MNRAS {\bf 431} (2013) 3282.

\bibitem{Bertsch1974}
G. F. Bertsch, Ann. Phys. {\bf 86} (1974) 138.

\bibitem{ks04}
M. Karlovini and L. Samuelsson, Classical and Quantum Gravity {\bf 21} (2004) 4531.

\bibitem{hs97}
L. Herrera and N. O. Santobs, Phys. Rep. {\bf 286} (1997) 53.

\bibitem{xty06}
R. X. Xu, D. J. Tao and Y. Yang, MNRAS, {\bf 373} (2006) L85.

\bibitem{psr68}
A. Hewish, J. Bell {\it et al.}, Nature {\bf 217} (1968) 709.

\bibitem{Gellmann1964}
M. Gell-Mann, Physics Letters {\bf 8} (1964) 214.

\bibitem{RS1975}
M. A. Ruderman and P. G. Sutherland, ApJ {\bf 196} (1975) 51.

\bibitem{Fowlers1977}
E. G. Flowers, M. A. Ruderman, J. F. Lee, P. G. Sutherland, W. Hillebrandt and E. Mueller, ApJ {\bf215} (1977) 291.

\bibitem{Lai2001}
D. Lai, Rev. Mod. Phys. {\bf 73} (2001) 629.

\bibitem{gm06}
J. Gil, G. Melikidze and B. Zhang,  ApJ {\bf 650} (2006) 1048.

\bibitem{YX2011}
J. W. Yu and R. X. Xu, MNRAS {\bf 414} (2011) 489.

\bibitem{XQZ1999}
R. X. Xu, G. J. Qiao and B. Zhang, ApJ {\bf 522} (1999) L109.

\bibitem{QLZ04}
G. J. Qiao, K. J. Lee, B. Zhang, R. X. Xu and H. G. Wang, ApJ {\bf 616} (2004) L127.

\bibitem{Romani1987}
R. W. Romani, ApJ {\bf 313} (1987) 718.

\bibitem{HL03}
W. C. G. Ho and D. Lai, MNRAS {\bf 338} (2003) 233.

\bibitem{TZD04}
R. Turolla, S. Zane and J. J. Drake, ApJ {\bf 603} (2004) 265.

\bibitem{Xu02}
R. X. Xu, ApJ {\bf 570} (2002) L65.

\bibitem{Sanwal2002}
D. Sanwal, G. G. Pavlov, V. E. Zavlin and M. Teter, ApJ {\bf 574} (2002) 61.

\bibitem{Mereghetti2002}
S. Mereghetti, A. De Luca, P. Caraveo, W. Becker, R. Mignani and G. F. Bignami, ApJ {\bf 581} (2002) 1280.

\bibitem{Bignami2003}
G. F. Bignami, P. A. Caraveo, A. De Luca and S. Mereghetti, Nature {\bf 423} (2003) 725.

\bibitem{XWQ2003}
R. X. Xu, H. G. Wang and G. J. Qiao, Chin. Phys. Lett. {\bf 20} (2003) 314.

\bibitem{GH2007}
E. V. Gotthelf and J. P. Halpern, ApJ {\bf 664} (2007) 35.

\bibitem{Xu2012}
R. X. Xu, S. I. Bastrukov, F. Weber, J. W. Yu and I. V. Molodtsova, Phys. Rev. D {\bf 85} (2012) 023008.

\bibitem{Zavlin1996}
V. E. Zavlin, G. G. Pavlov and Y. A. Shibanov, A\&A {\bf 315} (1996) 141.

\bibitem{zxz04}
X. L. Zhang, R. X. Xu and S. N. Zhang, A strange star with solid quark surface? in {\em Young neutron stars and their environments},
eds. F. Camilo and B. M. Gaensler (San Francisco, 2004) p.~303.

\bibitem{Zakharov10}
B. G. Zakharov, Phys. Lett. B {\bf 690} (2010) 250.

\bibitem{Paczynski1986}
B. Paczy\`{n}ski,  ApJ {\bf 308} (1986) 43.

\bibitem{Meszaros1998}
P. M\'{e}sz\'{a}ros, M. J. Rees and R. A. M. J. Wijers, ApJ {\bf 499} (1998) 301.

\bibitem{Ouyed05}
R. Ouyed, R. Rapp and C. Vogt, ApJ {\bf 632} (2005) 1001.

\bibitem{Paczynski2005}
B. Paczy\`{n}ski and P. Haensel, MNRAS {\bf 362} (2005) 4.

\bibitem{Cheng1996}
K. S. Cheng and Z. G. Dai, Phys. Rev. Lett. {\bf 77} (1996) 1210.

\bibitem{xu05mn}
R. X. Xu, MNRAS {\bf 356} (2005) 359.

\bibitem{Haensel1991}
P. Haensel, B. Paczy\`{n}ski and P. Amsterdamski, ApJ {\bf 375} (1991) 209.

\bibitem{CYX07}
A. B. Chen, T. H. Yu and R. X. Xu, ApJ {\bf 668} (2007) L55.

\bibitem{Dong16}
S. Dong, B. J. Shappee, J. L. Prieto, et al., Science {\bf 351} (2016) 257.

\bibitem{Stairs00}
I. H. Stairs, A. G. Lyne and S. L. Shemar, Nature {\bf 406} (2000) 484.

\bibitem{Xu07b}
R. X. Xu, Adv. Space Res. {\bf 40} (2007) 1453.

\bibitem{Alpar87}
M. A. Alpar, Phys. Rev. Lett. {\bf 58} (1987) 2152.

\bibitem{BHV90}
O. G. Benvenuto, J. E. Horvath and H. Vucetich, Phys. Rev. Lett. {\bf 64} (1990) 713.

\bibitem{z04}
A. Z. Zhou, R. X. Xu, X. J. Wu and N. Wang,  Astropart. Phys. {\bf 22} (2004) 73.

\bibitem{px07}
C. Peng and R. X. Xu, MNRAS {\bf 384} (2008) 1034.

\bibitem{zhou14}
E. P. Zhou, J. G. Lu, H. Tong and R. X. Xu, MNRAS {\bf 443} (2014) 2705.

\bibitem{tx11}
H. Tong and R. X. Xu, IJMP {\bf E20} (S2) (2011) 15.

\bibitem{lxz15}
M. X. Lin, R. X. Xu and B. Zhang, ApJ {\bf 799} (2015) 152.

\bibitem{Sergey09}
S. I. Bastrukov, G. T. Chen, H. K. Chang {\it et al.}, ApJ {\bf690} (2009) 998.

\bibitem{xl09}
R. X. Xu and E. W. Liang, Sci. Chin. Ser. G: Phys., Mech. Astron. {\bf 52} (2009) 315.

\bibitem{dlx11}
S. Dai, L. X. Li and R. X. Xu, Sci. Chin. Ser. G: Phys., Mech. Astron. {\bf 54} (2011) 1514.

\bibitem{LX09}
X. Y. Lai and R. X. Xu, MNRAS {\bf 398} (2009) L31.

\bibitem{Demorest12}
P. Demorest, T. Pennucci, S. Ransom, M. Roberts and J. Hessels, Nature {\bf 467} (2010) 1081.

\bibitem{Antoniadis13}
J. Antoniadis, P. C. C. Freire, N. Wex {\it et al.}, Science {\bf 340} (2013) 448.

\bibitem{Li2015}
Z. S. Li, Z. J. Qu, L. Chen {\it et al.}, ApJ {\bf 798} (2015) 56.

\bibitem{pavlov04}
G. G. Pavlov, D. Sanwal and M. A. Teter, in {\it Young Neutron Stars and Their Environments}, IAU Symposium no. {\bf 218} (2004) p.239.

\bibitem{PL09}
G. G. Pavlov and G. J. M. Luna, ApJ {\bf 703} (2009) 910.

\bibitem{HH09}
Wynn C. G. Ho and Craig O. Heinke, Nature {\bf 462} (2009) 71.

\bibitem{KK08}
M. H. van Kerkwijk and D. L. Kaplan, ApJ {\bf 673} (2008) L163.

\bibitem{HG10}
J. P. Halpern and E. V. Gotthelf, ApJ {\bf 709} (2010) 436.

\bibitem{LiXH05}
X. H. Li, F. J. Lu and T. P. Li, ApJ {\bf 628} (2005) 931.

\bibitem{1700}
R. X. Xu, RAA {\bf 14} (2014) 617.

\bibitem{Galloway02}
D. K. Galloway, J. L. Sokoloski and S. J. Kenyon, ApJ {\bf 580} (2002) 1065.

\bibitem{XW03}
R. X. Xu and F. Wu, Chin. Phys. Lett. {\bf 20} (2003) 80.

\bibitem{Madsen05}
J. Madsen, Phys. Rev. D {\bf 71} (2005) 014026.

\bibitem{LX10}
X. Y. Lai and R. X. Xu, J. Cos. \& Astropart. Phys. {\bf 5} (2010) 28.

\bibitem{LX16}
X. Y. Lai and R. X. Xu, RAA {\bf 16} (2016) in press (arXiv:1506.04172) .

\end{thebibliography}
\end{document}